\documentclass[11pt,a4paper]{article}
\pdfoutput=1
\usepackage{color}
\usepackage{bm}
\usepackage{amsmath,amssymb}
\usepackage{graphicx}

\numberwithin{figure}{section}
\numberwithin{equation}{section}

\usepackage[utf8]{inputenc}

\usepackage{hyperref} 
\hypersetup{colorlinks=true, citecolor=green, filecolor=black, linkcolor=blue, urlcolor=blue }
\usepackage{cite}

\newcommand\ie{\textit{i.e.}\ }
\newcommand\eg{\textit{e.g.}\ }

\newcommand{\derarg}{\left(\frac{D^2}{M^2}\right)}
\newcommand{\derbarg}{\left(\frac{\nabla^2}{M^2}\right)}
\newcommand{\w}{\underset{\rm weak}{=}}

\begin{document}

\begin{titlepage}
\begin{center}
{\huge \bf Cosmological backreaction in higher-derivative gravity expansions}
\end{center}

\begin{center}
{\bf Anthony W. H. Preston}
\end{center}

\begin{center}
{\it {Department} of Physics and Astronomy,  University of Southampton\\
Highfield, Southampton, SO17 1BJ, U.K.}\\
\vspace*{0.3cm}
{\tt  awhp1g12@soton.ac.uk}
\end{center}
\abstract{We calculate a general effective stress-energy tensor induced by cosmological inhomogeneity in effective theories of gravity where the action is Taylor-expandable in the Riemann tensor and covariant derivatives of the Riemann tensor.
This is of interest as an effective fluid that might provide an alternative to the cosmological constant, but it also applies to gravitational waves.
We use an adaptation of Green and Wald's weak-averaging framework, which averages over perturbations in the field equation where the perturbation length scales are small compared to the averaging scale.
In this adaptation, the length scale of the effective theory, $1/M$, is also taken to be small compared with the averaging scale. 
This ensures that the perturbation length scales remain in fixed proportion to the length scale of the effective theory as the cosmological averaging scale is taken to be large.
We find that backreaction from higher-derivative terms in the effective action can continue to be important in the late universe, given a source of sufficiently high-frequency metric perturbations.
This backreaction might also provide a window on exotic particle physics in the far ultraviolet.
} 
\end{titlepage}


\section{Introduction}

The standard cosmological concordance model is the $\Lambda$CDM model.
Assumed by this is the Cosmological Principle, that the universe is homogeneous and isotropic at sufficiently large distance scales and can thus the metric at large scales is accurately approximated by a Friedmann-Lema\^{i}tre-Robertson-Walker (FLRW) metric.
This is supported by astronomical observations, which indicate that the universe transitions to homogeneity at length scales of order 100 $h^{-1}$Mpc, see for example the analyses in \cite{Hogg:2004vw,Yadav:2005vv,Sarkar:2009iga,Scrimgeour:2012wt,Pandey:2015xea,Pandey:2015htc}.
Homogeneity has also been verified at a much larger volume scale of 14 $h^{-3}$Gpc in \cite{Laurent:2016eqo}.

The $\Lambda$CDM model considers a universe that consists of homogeneous fluids, which are the cosmological constant, cold dark matter, a small amount of baryonic matter and radiation, the latter being unimportant at late times.
At small distance scales, the universe is obviously not homogeneous. 
The density of the Earth is a factor of $10^{31}$ greater than the cosmological average and nucleons are a factor of $10^{46}$ more dense than the cosmological average. 
The standard cosmological model is a simplified picture in which it is assumed that these variations can be averaged out at large scales without introducing significant changes to the dynamics or expansion rate of the universe.

A problem with this view is that Einstein's field equations for General Relativity (GR) are non-linear, with the result that performing an averaging procedure on the equations does not merely return the same equations with an averaged metric, but rather includes extra terms that could be interpreted as additional effective fluids.
This effect is called cosmological backreaction.
The backreaction effects in Einstein's GR are typically considered to be small, radiation-like and unimportant in late-universe cosmology, as argued in \cite{Ishibashi:2005sj} and \cite{Behrend:2007mf}.
Much of the modern interest in backreaction comes from applying Buchert's averaging scheme, see for example \cite{Buchert:1995fz,Buchert:1999er,Buchert:2001sa,Visser:2015mur}.
Buchert's approach has also been applied to modified theories of gravity in \cite{Vitagliano:2009zy} and \cite{Jimenez:2013mwa}.
An extensive literature as accumulated on the r\^{o}le of inhomogeneity in cosmology, see for example \cite{Zalaletdinov:1992cg,Zalaletdinov:1996aj,Rasanen:2003fy,Alnes:2005rw,Coley:2006kp,Schwarz:2010px,Buchert:2011sx,Roy:2011za,Clarkson:2011zq,Roukema:2013cya,Clifton:2013vxa,Bentivegna:2015flc,Buchert:2015wwr,Sanghai:2016ucv}.

An elegant and rigorous framework for studying cosmological inhomogeneity has been proposed by Green and Wald in \cite{Green:2010qy}. 
They demonstrated their framework against specific examples in \cite{Green:2013yua}.
In this framework, an effective stress-energy tensor is calculated for the backreaction.
The method uses a generalization of Burnett's work on gravitational waves \cite{Burnett:1989gp} to the non-vacuum case where there exists a stress-energy tensor that satisfies the weak energy condition.
That in turn is a mathematically rigorous formulation of Isaacson's high-frequency approximation for the stress-energy tensor of gravitational waves from a distant source \cite{Isaacson:1967zz,Isaacson:1968zza}. 
The effective stress-energy tensor obtained, given Einstein's theory of gravity, has a vanishing trace, indicating that it is radiation-like.
Such a form for the backreaction cannot be important in late-universe cosmology.
This conclusion was contested in \cite{Buchert:2015iva}, the criticisms were responded to in \cite{Green:2015bma} and a simpler argument was provided in \cite{Green:2016cwo}.
If one accepts that the metric does converge under a suitable averaging procedure to a FLRW background at large distance scales, as argued in \cite{Green:2014aga}, this conclusion is rigorous, given Einstein's gravity theory.

Extending Einstein's GR to include higher-derivative terms can be motivated theoretically via constructing low-energy effective thoeries.
Effective theories can be constructed via Renormalization Group (RG) flows by integrating out high-energy modes down to some cutoff scale \cite{Polchinski:1983gv}.
The effective action will then appear as a series expansion in local operators with coefficients carrying mass dimensions as powers of the cutoff.
Locality is an essential feature, since RG is fundamentally underpinned by Kadanoff blocking \cite{Kadanoff:1966wm,Wilson:1971bg}.
A useful introduction can be found in \cite{Morris:1998da}.
When constructing an effective theory of gravity, it is usually required that the effective action is diffeomorphism-invariant, ensuring that its description is independent of our choice of spacetime coordinates.
In practice, this means that the effective action expansion is in the Riemann tensor and covariant derivative operators, with higher mass dimension operators being suppressed by higher powers of the cutoff.
A manifestly diffeomorphism-invariant Exact RG has recently been constructed for gravity at the classical level \cite{Morris:2016nda}.
In addition to the advantage of manifest diffeomorphism invariance, its background-independent construction allows for very easy implementation in this study of backreaction.
For an alternative exploration of background independence in the Exact RG for gravity, see \cite{Labus:2016lkh}.

Higher-derivative terms are also motivated phenomenologically for constructing cosmological models.
The Starobinsky $R^2$ term \cite{Davies:1977ze,Starobinsky:1980te}, provides a mechanism for early-universe inflation that is currently favoured by observations of the Cosmic Microwave Background, given a scale for $M\sim10^{13}$ GeV: see the results from WMAP \cite{Hinshaw:2012aka} and Planck \cite{Planck:2013jfk,Ade:2015lrj}.
The action for this can be written as
\begin{equation}\label{Staraction}
 S = \int d^4 x\sqrt{-g}\left[\frac{1}{16\pi G}\left(R+\frac{R^2}{6M^2}-2\Lambda\right)+\mathcal{L}_{\rm Matter}\right].
\end{equation}
The ``natural'' scale for quantum gravity is usually taken from Newton's gravitational constant in the natural units commonly used in particle physics to be of order $M \sim 10^{19}$ GeV, but a closer scale where interesting physics is anticipated is the Grand Unified Theory (GUT) scale of $10^{16}$ GeV, where a unification of the Standard Model gauge interactions is commonly expected.

This paper further generalizes work on backreaction in $R+R^2 /6M^2$ gravity developed in \cite{Preston:2014tua}, using an adaptation of the Green and Wald framework. 
An alternative approach that uses the Green and Wald framework under the approximation that the stress-energy tensor is set to its background form can be found in \cite{Saito:2012xa}.
Another study based on Green and Wald's formalism has considered the case where gravity is coupled to a massless scalar field \cite{Szybka:2015zca}.
The stress-energy tensor for gravitational waves in higher-derivative gravity has also been studied using Isaacson's formulation in \cite{Stein:2010pn} and \cite{Berry:2011pb}.
Both of these studies agree with this work that those Lagrangian terms that are of cubic order or higher in the Riemann tensor do not contribute to the effective stress-energy tensor, as will be discussed in Section \ref{FRrev}.
Both studies differ from this work in that they both fix a gauge rather than maintaining diffeomorphism invariance, which will be demonstrated for this work in Section \ref{DiffInvSect}.
Another difference is that \cite{Stein:2010pn} evaluates the stress-energy tensor at asymptotically-flat future null infinity, and \cite{Berry:2011pb} sets the background to Minkowski, whereas in this work we consider a general cosmological spacetime background.
The extension to higher-derivative gravity presents the challenge of incorporating the additional ``scalaron'' mass scale, $M$, into Green and Wald's existing framework.
As will be discussed also in this paper, the result found in \cite{Preston:2014tua} for the simple $R+R^2/6M^2$ theory was that the effective stress-energy tensor due to backreaction was no longer traceless, but rather possessed a negative-definite pure trace component.
This exciting result opened the possibility that cosmological backreaction in a theory of gravity with higher-derivative terms could effectively mimic a positive cosmological constant, offering an alternative to the standard $\Lambda$CDM model.

Since the scalaron mass would be expected to be large, corresponding to high-energy modes integrated out of a more fundamental theory of gravity, the effect of this term would be strongly suppressed unless the length scale of the perturbations were also of some very short distance scale.
To emphasise this point, let us fix a generous value for $M$ at $3\times 10^{13}$ GeV, as motivated by Starobinsky inflation, and let us consider the inhomogeneity as a single Fourier mode in the metric with wavelength similar to the radius of the Earth, \ie $L$ where $1/L\sim 3\times 10^{-23}$ GeV in natural units. An operator of the form $\nabla^2/M^2$ would then introduce a suppression by a factor of $\sim1/(ML)^2$, \ie 72 orders of magnitude. Larger length scales for the inhomogeneity, \eg galaxy clusters, result in even larger suppression of higher derivatives. To obtain a significant effect from the higher derivatives, we need to consider more high-frequency sources of inhomogeneity that would be related to high-energy particle physics, rather than the large structures ordinarily considered in astrophysics.
 
Suggested in the conclusions of \cite{Preston:2014tua} were two examples of exotic sources for such inhomogeneity.
The first suggestion was WIMPzillas \cite{Kolb:1998ki,Chung:2001cb}, however the dilution of the WIMPzilla density as the universe expands makes this suggestion phenomenologically unattractive.
The second, even more speculative, suggestion was that fluctuations in quantum spacetime might average to smooth classical perturbations at a length scale to which the effective gravity theory is sensitive.
Such an idea would risk problems with naturalness; it would, however, scale appropriately as the universe expands.
For an example of spontaneous breaking of translational symmetry in Planck-scale quantum gravity, see \cite{Bonanno:2013dja}.

A recent study proposed that the required inhomogeneity could be sourced from a vacuum that breaks translational symmetry in a non-Abelian gauge theory that is in a sector of particle physics disconnected from the Standard Model \cite{Evans:2015zwa}.
For a choice of $M\sim 10^{13}$ GeV, they concluded that an inhomogeneity wavelength at the electroweak scale would mimic a cosmological constant of the correct energy scale, which is $\sim 10^{-12}$ GeV. More specifically, the scaling of the effective vacuum energy in $R+R^2/6M^2$ gravity was estimated to be
\begin{equation}\label{EMSest}
 E_{\rm vac} \sim \frac{\Lambda^2_{\rm stripe}}{\sqrt{MM_{\rm Planck}}},
\end{equation}
where $E_{\rm vac}$ is the energy scale for the effective vacuum energy from backreaction, $\Lambda_{\rm stripe}\sim 100$ GeV is the energy scale that sets the amplitude and wavenumber of the translational symmetry violation and $M_{\rm Planck}\sim 10^{18}$ GeV is the reduced Planck mass.
The reason for disconnecting this sector from the Standard Model is to avoid introducing violations of Lorentz symmetry that would have already been observed, see for example \cite{Pospelov:2004fj,Pruttivarasin:2014pja}. Ordinarily, new physics of this kind would be inaccesible to experiment, however its effect on backreaction would be cosmologically observable. Put another way, one can constrain extensions of this kind to high-energy physics via the observable backreaction effect they would have if physically realized. An attraction of linking the inhomogeneity with the vacuum is that it would have the correct scaling as the universe expands.

In this paper, we will further generalize the Green and Wald framework to calculate the general form for the stress-energy tensor for a diffeomorphism-invariant higher-derivative gravity expansion.
To ensure that the averaging procedure converges in the weak limit, we will require that the field equations are Taylor-expandable in metric perturbations.
Together with diffeomorphism invariance, this translates into requiring that the action is Taylor-expandable in the Riemann tensor and covariant derivatives of the Riemann tensor.
This can be intuitively viewed as a locality requirement, as is reasonable for an averaging scheme.

This paper is structured as follows. Section \ref{Notation} outlines the notation conventions used in this paper. 
Section \ref{review} summarizes the features of the Green and Wald formalism and the extension to $f(R)$ models that are important for this paper.
In Section \ref{StressSection}, we calculate the generalization of the effective stress-energy tensor for local, manifestly diffeomorphism-invariant effective theories of gravity parametrized by a large mass scale.
The discussion and conclusions are given in Section \ref{CONCL}. Appendix \ref{appindiv} contains the results of applying the averaging procedure to individual field equation terms. Appendix \ref{apptrace} gives a consistency demonstration for the trace of the effective stress-energy tensor.
\section{Notation}\label{Notation}

We adopt Landau-Lifshitz spacelike sign conventions, (+,+,+), where the metric signature is mostly positive, a Ricci tensor defined as $R_{\mu\nu} := R^{\alpha}_{\ \mu\alpha\nu}$, and
\begin{equation}
 R^{\alpha}_{\ \beta\gamma\delta} = 2\partial_{[\gamma}\Gamma^{\alpha}_{\ \delta]\beta} + 2\Gamma^{\alpha}_{\ \lambda[\gamma}\Gamma^{\lambda}_{\ \delta]\beta}.
\end{equation}
We use the torsionless metric connection:
\begin{equation}
 \Gamma^{\alpha}_{\ \beta\gamma} = \frac{1}{2}g^{\alpha\lambda}\left(\partial_\beta g_{\gamma\lambda}+\partial_\gamma g_{\beta\lambda}-\partial_\lambda g_{\beta\gamma}\right).
\end{equation}
We will use a covariant derivative $D_\mu$ associated with the full metric $g_{\rho\sigma}$ such that when it acts on some tensor $T_{\alpha_1\cdots\alpha_m}^{\ \ \ \ \ \ \ \beta_1\cdots\beta_n}$, we have
\begin{equation}
 D_\mu T_{\alpha_1\cdots\alpha_m}^{\ \ \ \ \ \ \ \beta_1\cdots\beta_n} = \partial_\mu T_{\alpha_1\cdots\alpha_m}^{\ \ \ \ \ \ \ \beta_1\cdots\beta_n} - \sum_{i=1}^{m}\Gamma^{\lambda}_{\ \ \mu\alpha_i}T_{\alpha_1\cdots\lambda\cdots\alpha_m}^{\ \ \ \ \ \ \ \ \ \ \ \beta_1\cdots\beta_n} + \sum_{i=1}^{n}\Gamma^{\beta_i}_{\ \ \mu\lambda}T_{\alpha_1\cdots\alpha_m}^{\ \ \ \ \ \ \ \beta_1\cdots\lambda\cdots\beta_n}.
\end{equation}
A shorthand notation is used such that $D^2 = g^{\alpha\beta}D_\alpha D_\beta$.
We will also use a covariant derivative $\nabla_\mu$ that is associated with a background metric $g^{(0)}_{\mu\nu}$ such that
\begin{equation}
 D_\mu T_{\alpha_1\cdots\alpha_m}^{\ \ \ \ \ \ \ \beta_1\cdots\beta_n} = \nabla_\mu T_{\alpha_1\cdots\alpha_m}^{\ \ \ \ \ \ \ \beta_1\cdots\beta_n} - \sum_{i=1}^{m}C^{\lambda}_{\ \ \mu\alpha_i}T_{\alpha_1\cdots\lambda\cdots\alpha_m}^{\ \ \ \ \ \ \ \ \ \ \ \beta_1\cdots\beta_n} + \sum_{i=1}^{n}C^{\beta_i}_{\ \ \mu\lambda}T_{\alpha_1\cdots\alpha_m}^{\ \ \ \ \ \ \ \beta_1\cdots\lambda\cdots\beta_n},
\end{equation}
where $C^{\alpha}_{\ \beta\gamma}$ is the difference between full and background connections:
\begin{equation}\label{diffconnect}
 C^{\alpha}_{\ \beta\gamma} = \frac{1}{2}g^{\alpha\lambda}\left(\nabla_\beta h_{\gamma\lambda}+\nabla_\gamma h_{\beta\lambda}-\nabla_\lambda h_{\beta\gamma}\right).
\end{equation}
The full Ricci tensor can be split into background and perturbation parts by
\begin{equation}
 R_{\alpha\beta} = R^{(0)}_{\alpha\beta} - 2\nabla_{[\alpha}C^{\gamma}_{\ \gamma]\beta} + 2C^{\gamma}_{\ \beta[\alpha}C^{\delta}_{\ \delta]\gamma}.
\end{equation}
We will refer to the linear part of the perturbation to the Riemann tensor as
\begin{equation}
 R^{(1)}_{\alpha\beta\gamma\delta} := -2\nabla_{[\alpha|}\nabla_{[\gamma}h_{\delta]|\beta]}. 
\end{equation}
Using the notation that $\nabla^2=g^{(0)\alpha\beta}\nabla_{\alpha}\nabla_{\beta}$ and $h=g^{(0)\alpha\beta}h_{\alpha\beta}$, the linearized Ricci tensor is
\begin{equation}
 R^{(1)}_{\alpha\beta} := \frac{1}{2}\left(2\nabla_\lambda \nabla_{(\alpha}h_{\beta)}^{\ \ \lambda} - \nabla^2 h_{\alpha\beta} - \nabla_\alpha\nabla_\beta h\right).
\end{equation}
Finally, the linearized Ricci scalar is
\begin{equation}
 R^{(1)} := \nabla_\alpha \nabla_\beta h^{\alpha\beta} - \nabla^2 h.
\end{equation}
The linearized Riemann tensor and its contractions are invariant under linearized diffeomorphisms, given by
\begin{equation}
 \delta h_{\alpha\beta} = 2\nabla_{(\alpha}\xi_{\beta)}.
\end{equation}
Throughout this paper, we will be working in ``natural units'', as commonly used in high-energy physics, \ie $c=\hbar=1$.

\section{Weak-limit averaging method}\label{review}

In this section, we will summarize how the weak-averaging procedure developed by Green and Wald in \cite{Green:2010qy} has been adapted to higher-derivative gravity in \cite{Preston:2014tua}.
The weak-limit averaging procedure considers a one-parameter family of metrics whose inhomogeneity parameter, $\lambda$, is related to the wavelength of metric perturbations.
A tensor $A_{\alpha_1\cdots\alpha_n}(\lambda)$ converges in the weak limit to its ``average'' tensor, $B_{\alpha_1\cdots\alpha_n}$, if
\begin{equation}\label{genweaklim}
 \lim_{\lambda\to 0}\int d^4 x \sqrt{-g^{(0)}}f^{\alpha_1\cdots\alpha_n}A_{\alpha_1\cdots\alpha_n}(\lambda)=\int d^4 x\sqrt{-g^{(0)}}f^{\alpha_1\cdots\alpha_n}B_{\alpha_1\cdots\alpha_n}
\end{equation}
 for any smooth test field $f^{\alpha_1\cdots\alpha_n}$ of compact support.
The physical interpretation of $\lambda\to 0$ is that the averaging scale is much larger than the length scale of fluctuations, \ie the ratio of the wavelength of any perturbation mode to the averaging scale tends to zero as the averaging scale is taken to be large.
It will be convenient to denote equality under weak limit averaging as $\w$, such that if both $A_{\alpha_1\cdots\alpha_n}(\lambda)$ and another tensor, $C_{\alpha_1\cdots\alpha_n}(\lambda)$, converge in weak limit to $B_{\alpha_1\cdots\alpha_n}$, we can write
\begin{equation}
 A_{\alpha_1\cdots\alpha_n} \w B_{\alpha_1\cdots\alpha_n} \w C_{\alpha_1\cdots\alpha_n}.
\end{equation}
The metric, $g_{\mu\nu}(x,\lambda)$, can be separated into a $\lambda$-independent background metric, $g^{(0)}_{\mu\nu}(x)$, and a perturbation defined by
\begin{equation}
 h_{\mu\nu}(x,\lambda) := g_{\mu\nu}(x,\lambda) - g^{(0)}_{\mu\nu}(x).
\end{equation}
Note that we do not need to specify a particular choice of background.
The metric converges in the weak limit to the background metric, \ie the metric perturbation vanishes in weak limit.
More specifically, the metric perturbation is of $\mathcal{O}(\lambda)$, by which it is meant that $h_{\alpha\beta}(x,\lambda)$ is uniformly bounded by a constant times $\lambda$, for sufficiently small $\lambda$. Similarly, an $\mathcal{O}(\lambda^n)$ term is uniformly bounded by a constant times $\lambda^n$.
The choice of background metric that is best motivated by cosmology is the FLRW metric, but note that none of our calculations are specific to any particular choice of background, which can be chosen freely.
Applying a background covariant derivative to the metric perturbation lowers the order in $\lambda$ by one:
\begin{equation}\label{wavelength}
 \nabla_{\alpha_1}\cdots\nabla_{\alpha_n}h_{\beta\gamma} \sim \lambda^{1-n}.
\end{equation}
In this respect, we see that $\lambda$ is proportional to the length scales of perturbation modes.
However, this comes with a caveat.
Total derivatives, \ie derivatives that act on the entire tensor that we perform the weak limit averaging on, do not change the order in $\lambda$.
This can be seen by applying (\ref{genweaklim}) to a total derivative term and performing an integration by parts:
\begin{equation}
 \int d^4 x\sqrt{-g^{(0)}}f^{\lambda\alpha_1\cdots\alpha_n}\nabla_\lambda A_{\alpha_1\cdots\alpha_n}(\lambda) = -\int d^4x\sqrt{-g^{(0)}}\left(\nabla_\lambda f^{\lambda\alpha_1\cdots\alpha_n}\right)A_{\alpha_1\cdots\alpha_n}(\lambda).
\end{equation}
Since $f^{\alpha_1\cdots\alpha_n}$ is independent of $\lambda$, so are its derivatives, $\nabla_{\beta_1}\cdots\nabla_{\beta_n}f^{\alpha_1\cdots\alpha_n}$.
Thus we can see that the weak limit of a term consisting solely of total derivatives of $A_{\alpha_1\cdots\alpha_n}(\lambda)$ is of the same order in $\lambda$ as $A_{\alpha_1\cdots\alpha_n}(\lambda)$ itself.

This prescription will be applied to the field equations for a given gravity theory.
The field equation can be written as
\begin{equation}
 \mathcal{G}_{\mu\nu} := \frac{2\kappa}{\sqrt{-g}}\frac{\delta S_{\rm grav}}{\delta g^{\mu\nu}} = \kappa T_{\mu\nu},
\end{equation}
where $S_{\rm grav}$ is the gravitational part of the action, \ie the part constructed from a series expansion in the Riemann tensor and covariant derivatives of it, $\kappa=8\pi G$, and $T_{\mu\nu}$ is the stress-energy tensor.
We will wish to split $\mathcal{G}_{\mu\nu}$ into a background part, $\mathcal{G}_{\mu\nu}^{(0)}$ \ie the value of $\mathcal{G}_{\mu\nu}$ for $g_{\mu\nu}=g^{(0)}_{\mu\nu}$, and a perturbation part, $\delta\left[\mathcal{G}_{\mu\nu}\right]$.
The field equation is then written as
\begin{equation}\label{fieldpert}
 \mathcal{G}^{(0)}_{\mu\nu} + \delta\left[\mathcal{G}_{\mu\nu}\right] = \kappa T_{\mu\nu}.
\end{equation}
Supposing that the perturbation term is non-vanishing in the weak limit, we see that, moving it to the right hand side, we obtain an effective stress-energy tensor induced by inhomogeneity:
\begin{equation}
 \mathcal{G}^{(0)}_{\mu\nu} \w \kappa T_{\mu\nu}^{(0)} + \kappa t^{(0)}_{\mu\nu},
\end{equation}
where
\begin{equation}
 \delta\left[\mathcal{G}_{\mu\nu}\right] \w -\kappa t^{(0)}_{\mu\nu}.
\end{equation}
As discussed, linear terms in $h$ are of $\mathcal{O}(\lambda)$, because total derivatives do not change the order in $\lambda$.
Quadratic terms in $h$, on the other hand, can be of $\mathcal{O}(1)$, \ie being of the zeroth order in $\lambda$, they converge to finite values in the weak limit:
\begin{equation}
 h_{\rho\sigma}\nabla_{\alpha}\nabla_{\beta}h_{\mu\nu} \w - \nabla_{\alpha}h_{\rho\sigma}\nabla_{\beta}h_{\mu\nu} \sim \mathcal{O}(1).
\end{equation}
Terms of this form then provide non-vanishing contributions to the effective stress-energy tensor, $t_{\mu\nu}^{(0)}$. This is not an assertion that such terms are necessarily of the same order of magnitude as the background, rather that they are able to converge to give non-zero values in the limit where the averaging scale becomes large. This is subject to a constraint explored in the next section that can still result in some such terms vanishing, especially in Einstein gravity.

\subsection{Zero tensors}

There exists an additional constraint on $t_{\mu\nu}^{(0)}$ that is used in the papers by Isaacson \cite{Isaacson:1967zz,Isaacson:1968zza}, Burnett \cite{Burnett:1989gp} and Green and Wald \cite{Green:2010qy} for the Einstein gravity case.
This was referred to in \cite{Preston:2014tua} as the ``zero tensor''.
Since it will be used frequently here also, it is convenient to continue referring to it as the zero tensor.
In the Green and Wald paper, it is proven that
\begin{equation}\label{lemma}
 A(\lambda)B(\lambda) \w A(0)B(0),
\end{equation}
provided that $A(\lambda)$ is a smooth tensor field converging uniformly on compact sets to $A(0)$ and that $B(\lambda)$ is a non-negative smooth function converging to $B(0)$ in the weak limit.
Thus, provided that the stress-energy tensor $T_{\mu\nu}$ satisfies the weak energy condition, \ie given any timelike vector $t^{\alpha}(x,\lambda)$,
\begin{equation}
 T_{\alpha\beta}(x,\lambda)t^{\alpha}(x,\lambda)t^{\beta}(x,\lambda) \ge 0,
\end{equation}
it is also proven that
\begin{equation}\label{ZeroStress}
 h_{\rho\sigma}T_{\mu\nu} \w 0.
\end{equation}
Let us consider taking the field equation, as written in (\ref{fieldpert}), and multiplying by $h_{\rho\sigma}$:
\begin{equation}
 \underbrace{h_{\rho\sigma}\mathcal{G}^{(0)}_{\mu\nu}}_{\text{vanishes in weak limit}} + h_{\rho\sigma}\delta\left[\mathcal{G}_{\mu\nu}\right] = \underbrace{h_{\rho\sigma}\kappa T_{\mu\nu}}_{\text{vanishes in weak limit}}.
\end{equation}
Thus the form of the zero tensor is given by
\begin{equation}
 h_{\rho\sigma}\delta\left[\mathcal{G}_{\mu\nu}\right] \w 0,
\end{equation}
where, this time, it is the part of $\delta\left[\mathcal{G}_{\mu\nu}\right]$ that is linear in $h$ that gives us non-vanishing contributions.
This zero tensor constraint retains the information from the linear order in $h$, which the weak limit would otherwise discard.

\subsection{Application to Einstein gravity}

The action for Einstein's theory for gravity can be written as
\begin{equation}\label{EinAct}
 \int d^4 x \sqrt{-g}\frac{1}{2\kappa}\left(R -2\Lambda\right) + S_{\rm matter}.
\end{equation}
The field equation is
\begin{equation}
 R_{\mu\nu} -\frac{1}{2}g_{\mu\nu}R + \Lambda g_{\mu\nu} = \kappa T_{\mu\nu}.
\end{equation}
The cosmological constant term converges in the weak limit to its background value:
\begin{equation}\label{noccc}
 \Lambda g_{\mu\nu} \w \Lambda g^{(0)}_{\mu\nu}.
\end{equation}
This is clear because the metric converges in the weak limit to its background, and one is at liberty to multiply both sides by a constant.
Weak limits of the other two terms are given in Appendix \ref{appindiv}, along with the weak limits of more terms that appear in higher-derivative gravity.
The form of the effective stress-energy tensor, before applying the zero tensor constraint, is then
\begin{equation}\label{EinsteinStress}
 \kappa t^{E}_{\mu\nu} \w \frac{1}{2}h^{\alpha\beta}R_{\mu\alpha\nu\beta}^{(1)}+\frac{3}{4}h_{\mu\nu}R^{(1)} - R^{(1)}_{\alpha(\mu}h_{\nu)}^{\ \ \alpha} - \frac{1}{8}g_{\mu\nu}^{(0)}\left(hR^{(1)}+2h^{\alpha\beta}R_{\alpha\beta}^{(1)}\right).
\end{equation}
The zero tensor can be powerfully expressed for Einstein gravity as
\begin{equation}
 R_{\alpha\beta}^{(1)}h_{\gamma\delta} \w 0.
\end{equation}
This immediately simplifies (\ref{EinsteinStress}) to
\begin{equation}
 \kappa t^{E}_{\mu\nu} \w \frac{1}{2}h^{\alpha\beta}R_{\mu\alpha\nu\beta}^{(1)},
\end{equation}
which is then traceless under the constraint from the zero tensor, implying that the backreaction is radiation-like.
The implication of this is that cosmological backreaction is unable to account for accelerating expansion and is not important in late-time cosmology, which was the conclusion in \cite{Green:2010qy}.
\subsection{Extension to higher derivatives}\label{FRrev}

We will review the application of weak-limit averaging to a Taylor-expandable $f(R)$ model whose action has its own length scale, $M$.
This procedure was peformed explicitly for the $R+R^2 /6M^2$ case in \cite{Preston:2014tua}.
As we will see, the results are already general for the complete local $f(R)$ expansion.
Consider the following action:
\begin{equation}
 S = \int d^4 x\sqrt{-g}\left(f(R)-2\Lambda\right) + S_{\rm matter},
\end{equation}
where
\begin{equation}
 f(R) = R + \frac{R^2}{6M^2} + {\rm const}\times \frac{R^3}{M^4} + \cdots.
\end{equation}
The inclusion of a new length scale, $1/M$, presents us with the challenge of how to rigorously incorporate it. 
Recall that the limit $\lambda\to 0$ corresponds to choosing a cosmological averaging length scale that is much greater than the length scale of perturbations.
The na\"{i}ve suggestion that $M$ should be independent of $\lambda$ presents us with problems.
Firstly, consider the ratio of the scale $1/M$ and the perturbation scale, which goes like $\lambda$, see (\ref{wavelength}).
The ratio between these two scales should be fixed physically, since it is the cosmological averaging scale that we are tending to be large.
Leaving $1/M$ independent of $\lambda$ would correspond to taking the perturbation length scale to be much smaller than $1/M$ in the limit where we take the cosmological averaging scale to be large.
This would be outside the validity of the effective theory. 
The perturbations would have a length scale that is much shorter than $1/M$, which is the scale to which high-frequency modes in the more fundamental theory have been integrated out.
This would manifest as each higher-derivative term in the action giving increasingly divergent field equation contributions in the weak limit.
The na\"{i}ve suggestion that the latter problem could be fixed by rescaling to $h\sim\lambda^2$ would fail once one also allows the action to use explicit covariant derivative operators, \eg in Lagrangian terms like $RD^2 R/M^4$.

All of these problems are resolved by setting $M\sim\lambda^{-l}$ where $l=1$.
In particular, $l=1$ is the only scaling that keeps a fixed ratio of the perturbation wavelength and the scale $1/M$ in the weak limit.
Setting $l<1$ would result in divergences in the field equation caused by higher derivatives.
Setting $l>1$ would cause all contributions from Lagrangian terms with more than two derivatives to vanish, trivially leaving us with only the Einstein gravity contributions.
For the reasons discussed, we set $l=1$, with the result that
\begin{equation}\label{soleform}
 \frac{1}{M^{k-2}}h_{\mu\nu}\nabla_{\alpha_1}\cdots\nabla_{\alpha_k}h_{\rho\sigma} \w \mathcal{O}(1),
\end{equation}
and other non-zero orders in $h$ vanish in the weak limit.

The field equation for $f(R)$ gravity can be written as
\begin{equation}
 \left(R_{\mu\nu} -D_\mu D_\nu + g_{\mu\nu}D^2 \right)f'(R) -\frac{1}{2}g_{\mu\nu}f(R) + \Lambda g_{\mu\nu} = \kappa T_{\mu\nu}.
\end{equation}
All contributions to $t_{\mu\nu}^{(0)}$ from the $R^3$ and higher action terms vanish in the weak limit.
To see this, note that, for $m\ge0$ and $n>0$,
\begin{equation}\label{vanishing}
 \frac{1}{M^{2(m+n)}}h_{\alpha\beta}\nabla_\gamma\nabla_\delta R^{(1)n}R^{(0)m}\sim \lambda^{2m+n-1},
\end{equation}
thus our contributions from higher derivative terms are non-vanishing for $n=1$ and $m=0$.
Alternatively, note that (\ref{soleform}) tells us that the only non-vanishing perturbation contributions (at any order in derivatives) are from the quadratic order in the metric perturbation where all of the covariant derivative operators are acting on an instance of the metric perturbation not as a total derivative.
More generally, any field equation contribution from a Lagrangian term at cubic order or higher in the Riemann tensor vanishes in weak limit.
The $f(R)$ field equation expands up to $\mathcal{O}(R^2)$ as
\begin{equation}\label{Starofield}
 R_{\mu\nu} + \Lambda g_{\mu\nu} -\frac{1}{2}g_{\mu\nu}\left(R+\frac{R^2}{6M^2}\right)+\frac{1}{3M^2}\left(RR_{\mu\nu}-D_\mu D_\nu R + g_{\mu\nu}D^2 R\right)\cdots = \kappa T_{\mu\nu}
\end{equation}
Once again, upon taking a weak limit, we find that the Einstein (2-derivative) part of $\kappa t^{(0)}_{\mu\nu}$ is given by (\ref{EinsteinStress}).
The Starobinsky (4-derivative) part of the field equation also gives a contribution to $\kappa t^{(0)}_{\mu\nu}$, which we will denote by $\kappa t^{S}_{\mu\nu}$.

As before, the field equation in the weak limit can be written as
\begin{equation}
 R_{\mu\nu}^{(0)} -\frac{1}{2}g_{\mu\nu}^{(0)}R^{(0)} \w \kappa T_{\mu\nu} + \kappa t^{(0)}_{\mu\nu},
\end{equation}
where the higher-derivative contributions to $\mathcal{G}^{(0)}_{\mu\nu}$ vanish in the weak limit because of the scaling of $M$ with $\lambda$, \ie the higher-derivative background terms become unimportant in the limit where the averaging scale is very large.
However, this is not true of $t_{\mu\nu}^{S}$, which does not vanish in the weak limit.
For this reason, backreaction provides a rare opportunity for these terms to have an influence on cosmology at very large scales.

Weak limits of the individual field equation terms in (\ref{Starofield}) are given in Appendix \ref{appindiv}.
Putting these ingredients together, we find the contributions to the effective stress-energy tensor from the Starobinsky parts of the field equation:
\begin{equation}\label{PureStarostress}
 \kappa t_{\mu\nu}^{S} \w \frac{R^{(1)}}{3M^2}\left(\frac{1}{2}g_{\mu\nu}^{(0)}\nabla^2 h- \nabla^2 h_{\mu\nu}+\frac{1}{2}\nabla_\mu \nabla_\nu h + \frac{1}{4}g_{\mu\nu}^{(0)}R^{(1)}\right).
\end{equation}
There are two particularly useful forms of the zero tensor for this theory:
\begin{equation}\label{Szero1}
 h_{\alpha\beta}R^{(1)}_{\gamma\delta} -\frac{g_{\gamma\delta}^{(0)}}{6M^2}R^{(1)}\nabla^2 h_{\alpha\beta}-\frac{1}{3M^2}\left(\nabla_\gamma\nabla_\delta h_{\alpha\beta}\right)R^{(1)} \w 0
\end{equation}
and
\begin{equation}\label{Szero2}
 h_{\alpha\beta}R^{(1)}_{\gamma\delta} - \frac{g^{(0)}_{\gamma\delta}}{6}h_{\alpha\beta}R^{(1)}-\frac{1}{3M^2}\left(\nabla_\gamma \nabla_\delta h_{\alpha\beta}\right)R^{(1)} \w 0.
\end{equation}
These are both related via the trace over $\gamma\delta$:
\begin{equation}
 R^{(1)}\left(1-\frac{\nabla^2}{M^2}\right)h_{\alpha\beta} \w 0.
\end{equation}
Equation (\ref{Szero1}), having only a single very generic 2-derivative term, is useful for converting 2-derivative terms into 4-derivative terms.
This is especially true when considering the trace of the effective stress-energy tensor, $t^{(0)}$, where, as we will see, it is possible to rewrite every term in 4-derivative form.
This is because $t^{(0)}$ would vanish if not for the 4-derivative extension of the action, \ie the Einstein gravity case has a radiation-like backreaction.
Equation (\ref{Szero2}), conversely, having only a single very generic 4-derivative term, is useful for converting 4-derivative terms into 2-derivative terms.
In fact, (\ref{Szero2}) can be used to rephrase $t^{(0)}_{\mu\nu}$ entirely in terms of 2-derivative terms.
This is because, in a pure $R^2$ theory, $t_{\mu\nu}^{(0)}$ vanishes completely.
Thus we see that the r\^{o}le of the zero tensor is not as powerful as in the Einstein-gravity case, but it is still able to rewrite $t_{\mu\nu}^{(0)}$ into more convenient forms.
The 2-derivative form of the Starobinsky part of $t_{\mu\nu}^{(0)}$ is
\begin{equation}
 \kappa t^{S}_{\mu\nu} \w \frac{1}{2}hR_{\mu\nu}^{(1)} + \frac{g^{(0)}_{\mu\nu}}{4}h^{\alpha\beta}R_{\alpha\beta}^{(1)}-\frac{1}{3}h_{\mu\nu}R^{(1)}-\frac{g_{\mu\nu}^{(0)}}{24}hR^{(1)}.
\end{equation}
Putting the pieces together, the effective stress-energy tensor for this simple model in 2-derivative form is
\begin{equation}
 \kappa t^{(0)}_{\mu\nu} \w \frac{1}{2}h^{\alpha\beta}R^{(1)}_{\mu\alpha\nu\beta}-R^{(1)}_{\alpha(\mu}h_{\nu)}^{\ \ \alpha}+\frac{5}{12}h_{\mu\nu}R^{(1)}+\frac{1}{2}hR_{\mu\nu}^{(1)}-\frac{1}{6}g_{\mu\nu}^{(0)}hR^{(1)},
\end{equation}
for which the trace is
\begin{equation}
 \kappa t^{(0)} \w \frac{1}{4}hR^{(1)}-\frac{1}{2}h^{\alpha\beta}R_{\alpha\beta}^{(1)}.
\end{equation}
The 4-derivative form of the trace is
\begin{equation}\label{StarStressTrace}
 \kappa t^{(0)} \w -\frac{R^{(1)2}}{6M^2}.
\end{equation}
from this, we can immediately see that the inclusion of the Starobinsky $R^2$ term has given us a form for $t_{\mu\nu}^{(0)}$ that is not purely radiation-like.
Splitting the stress-energy tensor into its traceless and pure trace components, \ie
\begin{equation}
 t^{(0)}_{\mu\nu} = \underbrace{t^{(0)}_{\mu\nu} - \frac{1}{4}g_{\mu\nu}^{(0)}t^{(0)}}_{\text{traceless}}+\underbrace{\frac{1}{4}g^{(0)}_{\mu\nu}t^{(0)}}_{\text{pure trace}},
\end{equation}
we see that the pure trace component of $t_{\mu\nu}^{(0)}$ has the correct sign and order in $\lambda$ to mimic a positive cosmological constant, given an appropriate cosmological context. 
The diffeomorphism invariance of this construction was demonstrated in \cite{Preston:2014tua}, but we will postpone the derivation here until Section \ref{DiffInvSect}, where we demonstrate diffeomorphism invariance for the fully generalized case.
Also discussed in \cite{Preston:2014tua} was the equivalent scalar-tensor construction, which is formed via a Legendre transform, $f(R)=\phi R - V(\phi)$, where we would require that $V(\phi)$ is Taylor-expandable in $\phi$ so that it converges in the weak limit. 
Perturbations in $\phi$ would only contribute to $t_{\mu\nu}^{(0)}$ up to their quadratic order, beyond which they would be suppressed by $\lambda$. 

Na\"{i}vely, one might also think to scale the cosmological constant dimensionally, \ie $\Lambda\sim\lambda^{-2}$.
Na\"{i}vely again, one might think that the cosmological constant would then give an $\mathcal{O}(1)$ contribution to the zero tensor of the form $h_{\rho\sigma}h_{\mu\nu}\Lambda$.
However, this would give us a $\lambda^{-2}$ divergence in the field equation. 
For the sake of the convergence of the weak-limit and the overall consistency of the formalism, we must leave $\Lambda$ constant.
This does not introduce the pathology of changing the ratio of physical scales in the weak limit because the cosmological constant does not mix with other scales in the field equation.
To see this, note that (\ref{noccc}) does not have any implications for the relative sizes of $\Lambda^{-1/2}$, the perturbation wavelength and $1/M$.
This is unlike the higher-derivative terms, which effectively take a ratio of $1/M$ to the perturbation wavelength via the $M$-suppressed higher derivative operators.
A cosmological constant term that converges in the weak limit, as given in (\ref{noccc}), does not contribute to the zero tensor.
To see this, note that a positive cosmological constant term satisfies the weak energy condition and therefore (\ref{ZeroStress}) applies here:
\begin{equation}
 h_{\rho\sigma}g_{\mu\nu}\Lambda \w 0.
\end{equation}
This result is also true for a negative cosmological constant.
Thus the cosmological constant does not have any influence on the backreaction.

\section{Generalized backreaction in higher-derivative gravity}\label{StressSection}

\subsection{Contributing action terms}\label{contrib}

As discussed in Section \ref{FRrev}, especially via equations (\ref{soleform}) and (\ref{vanishing}), the only local action terms that give non-vanishing contributions to $t_{\mu\nu}^{(0)}$ are those that are linear or quadratic in the Riemann tensor.
Since the Gauss-Bonnet term is a topological invariant in four dimensions,
\begin{equation}\label{Gauss-Bonnet}
 \frac{\delta}{\delta g^{\mu\nu}}\int d^4x \sqrt{-g}\left(R_{\alpha\beta\gamma\delta}R^{\alpha\beta\gamma\delta} -4 R_{\alpha\beta}R^{\alpha\beta} + R^2\right) = 0,
\end{equation}
we can write a local expansion in the Riemann tensor up to the quadratic order as
\begin{equation}
 S = \int d^4 x \sqrt{-g}\frac{1}{2\kappa}\left(-2\Lambda + R + \frac{a}{M^2}R_{\mu\nu}R^{\mu\nu} + \frac{b}{M^2}R^2 + \cdots\right)+S_{\rm matter}.
\end{equation}

We can further generalize this action while maintaining diffeomorphism invariance and a suitable notion of locality by also introducing explicit covariant derivative operators in the action, again in a Taylor-expandable structure.
The first instance in the action where we can introduce these covariant derivative operators in a non-trivial way is at the quadratic order in the Riemman tensor \eg in a Lagrangian term like $R_{\mu\nu}a\derarg R^{\mu\nu}$.
Since, as already discussed, our expression for $t^{(0)}_{\mu\nu}$ is unaffected by cubic and higher terms in the Riemann tensor, we only need to consider explicit covariant derivatives in terms at quadratic order in the Riemann tensor.
We are free to rearrange the order of these covariant derivatives, since the commutators only introduce corrections at higher order in the Riemann tensor, \eg:
\begin{equation}
 \left[D_{\mu},D_{\nu}\right]v_{\alpha} = R^{\lambda}_{\ \alpha\nu\mu}v_{\lambda}.
\end{equation}

Na\"{i}vely, one would expect that there would exist a large number of independent index structures at quadratic order in the Riemann tensor.
However, using the (anti-)symmetry properties of the Riemann tensor and the (second) Bianchi identity:
 \begin{equation}
  D_{\lambda}R_{\alpha\beta\gamma\delta} + D_{\gamma}R_{\alpha\beta\delta\lambda} + D_{\delta}R_{\alpha\beta\lambda\gamma} = 0,
 \end{equation}
a particularly useful specialization of which is
\begin{equation}
 g^{\alpha\beta}D_{\alpha}R_{\beta\gamma} = \frac{1}{2}D_{\gamma}R,
\end{equation}
we are able to rearrange all possible index structures, up to the quadratic order in the Riemann tensor, to
\begin{eqnarray}\label{unsimpact}
 S & = & \int d^4x \sqrt{-g}\frac{1}{2\kappa}\left(R -2\Lambda + \frac{1}{M^2}R_{\alpha\beta}a\derarg R^{\alpha\beta} + \frac{1}{M^2}Rb\derarg R \right. \nonumber \\ && \left.+ \frac{1}{M^2}R_{\alpha\beta\gamma\delta}c\derarg R^{\alpha\beta\gamma\delta}+ \cdots\right) + S_{\rm matter}.
\end{eqnarray}
However, not all of these terms give independent contributions to $t_{\mu\nu}^{(0)}$.
As already discussed, the three quadratic terms without extra covariant derivative operators are related via the Gauss-Bonnet topological invariant (\ref{Gauss-Bonnet}). 
Although the introduction of the covariant derivatives breaks the topological invariance, the effect of this is not apparent in the weak limit, \ie there are no new corrections of the form (\ref{soleform}).
To see this, one can consult equations (\ref{derivA}) and (\ref{derivB}), noting that all the terms that contribute in the weak limit are of the same structure as in the case of keeping $a$ and $b$ as constant numbers, except for the final terms with $a'$ and $b'$.
Even these extra terms cancel if we choose the Gauss-Bonnet structure because 
 \begin{equation}\label{GB2}
  \frac{1}{M^4}\left(R_{\alpha\beta\gamma\delta}^{(1)}a' \nabla_\mu \nabla_\nu R^{\alpha\beta\gamma\delta(1)} -4R_{\alpha\beta}^{(1)}a' \nabla_\mu \nabla_\nu R^{\alpha\beta(1)} + R^{(1)}a'\nabla_\mu \nabla_\nu R^{(1)}\right) \w 0.
 \end{equation}
Thus the topological invariance of the Gauss-Bonnet term given in (\ref{Gauss-Bonnet}) is effectively inherited by its higher-derivative generalization in the weak limit. 
More precisely, we mean that
\begin{equation}\label{genGauss-Bonnet}
 \frac{\delta}{\delta g^{\mu\nu}}\int d^4x \sqrt{-g}\frac{1}{M^2}\left(R_{\alpha\beta\gamma\delta}a\derarg R^{\alpha\beta\gamma\delta} -4 R_{\alpha\beta}a\derarg R^{\alpha\beta} + Ra\derarg R\right) \w 0.
\end{equation}

For this reason, when calculating $t_{\mu\nu}^{(0)}$, we lose nothing by rearranging the three quadratic terms in the action to a basis of just two of them.
Finally, we are left with a concise, closed form for the contributing terms in the effective action that still gives us a completely general result for the form of $t_{\mu\nu}^{(0)}$:
\begin{equation}\label{TotalAction}
 S = \int d^4x \sqrt{-g}\frac{1}{2\kappa}\left(R -2\Lambda + \frac{1}{M^2}R_{\alpha\beta}a\derarg R^{\alpha\beta} + \frac{1}{M^2}Rb\derarg R + \cdots\right) + S_{\rm matter}.
\end{equation}

\subsection{Calculation of the effective stress-energy tensor}

Let us consider the action given in (\ref{TotalAction}).
To obtain the field equation, the ingredients we need are
\begin{eqnarray}\label{derivA}
  \frac{\delta}{\delta g^{\mu\nu}}\int d^4x \sqrt{-g}R_{\alpha\beta}aR^{\alpha\beta} & = & \sqrt{-g}\left(-\frac{1}{2}g_{\mu\nu}R_{\alpha\beta}aR^{\alpha\beta}+2R_{\mu\alpha}aR_{\nu}^{\ \alpha} \right.\nonumber \\ & &\left. +D^2 aR_{\mu\nu}+\frac{1}{2}g_{\mu\nu}D^2 aR-2D_\alpha D_{(\mu} aR_{\nu)}^{\ \ \alpha} \right.\nonumber \\ & &\left. + \frac{1}{M^2}R_{\alpha\beta}D_{\mu}D_{\nu}a'R^{\alpha\beta}+\cdots\right),
\end{eqnarray}
\begin{eqnarray}\label{derivB}
  \frac{\delta}{\delta g^{\mu\nu}}\int d^4x \sqrt{-g}RbR & = & \sqrt{-g}\left(-\frac{1}{2}g_{\mu\nu}RbR + 2RbR_{\mu\nu}-2D_{\mu}D_{\nu}bR + 2g_{\mu\nu}D^2 bR \right. \nonumber \\ & & \left. + \frac{1}{M^2}R D_{\mu}D_{\nu}b'R+\cdots\right),
\end{eqnarray}
where $a'$ and $b'$ are the derivatives of $a$ and $b$ with respect to their arguments, and we have neglected terms that vanish entirely in the weak limit.
The vanishing terms come from varying with respect to the inverse metric the connections from applying the explicit covariant derivative operators that appear in $a$ and $b$.
As before, we can take the weak limit to find the effective stress-energy tensor from metric perturbations.
In all of the terms above, the functions $a\derarg$ and $b\derarg$ converge simply in the weak limit to $a\derbarg$ and $b\derbarg$, respectively.
This is because they do not appear as total derivatives in any of the field equation terms.
By this, we mean, for example,
\begin{equation}
 \frac{1}{M^2}Rb\derarg R_{\mu\nu} \w \frac{1}{M^2} R^{(1)}b\derbarg R^{(1)}_{\mu\nu}.
\end{equation}
We have already calculated the weak limits of the field equation terms that are also found in $R+R^2 /6M^2$ gravity.
The generalized action now gives us new terms whose forms did not appear in that model, these are also given in Appendix \ref{appindiv}.
Putting together all the pieces, the effective stress-energy tensor in the weak limit is
\begin{eqnarray}\label{totalstress}
 \kappa t^{(0)}_{\mu\nu} & = & \frac{1}{2}h^{\alpha\beta}R^{(1)}_{\mu\alpha\nu\beta} + \frac{3}{4}h_{\mu\nu}R^{(1)}-R^{(1)}_{\alpha(\mu}h_{\nu)}^{\ \ \alpha}-\frac{1}{8}g^{(0)}_{\mu\nu}\left(hR^{(1)}+2h^{\alpha\beta}R^{(1)}_{\alpha\beta}\right) \nonumber \\ &&
+\frac{1}{M^2}\left(\frac{1}{2}g^{(0)}_{\mu\nu}R^{(1)}bR^{(1)}+h\nabla_{\mu}\nabla_{\nu}bR^{(1)}-2h_{\mu\nu}\nabla^2 bR^{(1)}+g^{(0)}_{\mu\nu}h\nabla^2 bR^{(1)} \right. \nonumber \\ &&
+ \frac{1}{2}g_{\mu\nu}^{(0)}R^{(1)}_{\alpha\beta}aR^{(1)\alpha\beta}-2R_{\mu\alpha}^{(1)}aR_{\nu}^{(0)\alpha}+ \frac{1}{2}h\nabla^2 a R_{\mu\nu}^{(1)}-2h^{\alpha}_{\ (\mu}\nabla^2aR_{\nu)\alpha}^{(1)}+h^{\alpha\beta}\nabla_\mu \nabla_\nu aR_{\alpha\beta}^{(1)} \nonumber \\ &&
+ h^\alpha_{\ (\mu}\nabla_{\nu)}\nabla_\alpha a R^{(1)} + \frac{1}{4}g_{\mu\nu}^{(0)}h\nabla^2 a R^{(1)}-\frac{1}{2}h\nabla_\mu \nabla_\nu a R^{(1)} - \frac{1}{2}h_{\mu\nu}\nabla^2 aR^{(1)}\nonumber \\ &&
\left. -R^{(1)}\frac{\nabla_\mu \nabla_\nu}{M^2}b'R^{(1)} -R^{(1)}_{\alpha\beta}\frac{\nabla_\mu \nabla_\nu}{M^2}a'R^{(1)\alpha\beta}\right).
\end{eqnarray}
To gain more insight into this stress-energy tensor, we derive the corresponding zero tensor in the next section.

\subsection{The zero tensor}

As before, we derive the zero tensor by multiplying the field equation by $h_{\rho\sigma}$ and taking a weak limit.
We get
\begin{eqnarray}\label{rawzero}
 0 & \w & h_{\rho\sigma}R^{(1)}_{\mu\nu} - \frac{1}{2}g^{(0)}_{\mu\nu}h_{\rho\sigma}R^{(1)} + \frac{1}{M^2}\left(-2h_{\rho\sigma}\nabla_\mu \nabla_\nu bR^{(1)} + 2g_{\mu\nu}^{(0)}h_{\rho\sigma}\nabla^2 bR^{(1)} \right. \nonumber \\ &&
\left. h_{\rho\sigma}\nabla^2 aR_{\mu\nu}^{(1)}+\frac{1}{2}g_{\mu\nu}^{(0)}h_{\rho\sigma}\nabla^2 aR^{(1)} - h_{\rho\sigma}\nabla_\mu \nabla_\nu aR^{(1)}\right).
\end{eqnarray}
Taking the trace over $\mu$ and $\nu$, we get
\begin{equation}
 h_{\rho\sigma}R^{(1)} \w \frac{2}{M^2}h_{\rho\sigma}(a+3b)\nabla^2 R^{(1)}.
\end{equation}
Substituting this back into (\ref{rawzero}) allows us to write the zero tensor in other forms, the most useful of which is
\begin{equation}\label{genzero}
 h_{\rho\sigma}R^{(1)}_{\mu\nu} \w \frac{1}{M^2}\left(h_{\rho\sigma}\nabla_\mu \nabla_\nu (a+2b)R^{(1)} + \frac{1}{2}g_{\mu\nu}^{(0)}h_{\rho\sigma}\nabla^2 (a+2b)R^{(1)}-h_{\rho\sigma}\nabla^2 a R_{\mu\nu}^{(1)}\right).
\end{equation}
A useful contraction of this form is
\begin{equation}
 h_{\alpha\beta}R^{(1)\alpha\beta} \w \frac{1}{M^2}\left(h^{\alpha\beta}\nabla_\alpha \nabla_\beta (a+2b)R^{(1)}+\frac{1}{2}h\nabla^2 (a+2b)R^{(1)}-h^{\alpha\beta}\nabla^2 a R^{(1)}_{\alpha\beta}\right).
\end{equation}

\subsection{Trace of the effective stress-energy tensor}

In this section, we will take the trace of the field equation and perform the weak limit to find
the trace of the effective stress-energy tensor. This is equivalent to taking the trace of (\ref{totalstress}), as
demonstrated in Appendix \ref{apptrace}.
Taking the trace of the field equation, we get
\begin{equation}
 -R + \frac{2}{M^2}D^2 (a+3b)R + \frac{1}{M^4}\left(R_{\alpha\beta}D^2 a'R^{\alpha\beta} + RD^2 b'R\right)+\cdots= \kappa T.
\end{equation}
Noting that
\begin{eqnarray}
 -R^{(2)} & \w & \frac{1}{4}hR^{(1)}+\frac{1}{2}h^{\alpha\beta}R^{(1)}_{\alpha\beta} \nonumber \\ & \w &
\frac{1}{2M^2}\left(h^{\alpha\beta}\nabla_\alpha \nabla_\beta (a+2b)R^{(1)}+\frac{1}{2}h\nabla^2 (3a+8b)R^{(1)}-h^{\alpha\beta}\nabla^2 a R^{(1)}_{\alpha\beta}\right),
\end{eqnarray}
and
\begin{equation}
 \frac{1}{M^2}\delta\left[D^2 (a+3b)R\right] \w -\frac{1}{2M^2}h\nabla^2 (a+3b)R^{(1)},
\end{equation}
we can derive the form of $\kappa t^{(0)}$ in the weak limit to be
\begin{eqnarray}\label{intermtrace}
 \kappa t^{(0)} & \w & -\frac{1}{2M^2}\left(h^{\alpha\beta}\nabla_\alpha \nabla_\beta (a+2b)R^{(1)}-h\nabla^2 (\frac{a}{2}+2b)R^{(1)}-h^{\alpha\beta}\nabla^2 aR^{(1)}_{\alpha\beta}\right. \nonumber \\ && 
\left. +2R^{(1)}_{\alpha\beta}a'\frac{\nabla^2}{M^2}R^{(1)\alpha\beta} + 2R^{(1)}b'\frac{\nabla^2}{M^2}R^{(1)}\right).
\end{eqnarray}
A more elegant way of writing this can be found by separating the first three terms into two expressions. Firstly,
\begin{equation}
 \frac{1}{M^2}\left(h^{\alpha\beta}\nabla_{\alpha}\nabla_{\beta}(a+2b)R^{(1)} - h\nabla^2 (a+2b)R^{(1)}\right) \w \frac{1}{M^2}R^{(1)}(a+2b)R^{(1)},
\end{equation}
but also, less obviously,
\begin{equation}
 \frac{1}{M^2}\left(\frac{1}{2}h\nabla^2 aR^{(1)}-h^{\alpha\beta}\nabla^2 a R^{(1)}_{\alpha\beta}\right) \w \frac{1}{M^2}\left(2R^{(1)}_{\alpha\beta}aR^{(1)\alpha\beta} - R^{(1)}aR^{(1)}\right).
\end{equation}
Now we can simplify (\ref{intermtrace}) to a more elegant, more clearly diffeomorphism-invariant form:
\begin{equation}\label{totalstresstrace}
 \kappa t^{(0)} \w  -\frac{1}{M^2}\left(R^{(1)}_{\alpha\beta}\left(a+ a'\frac{\nabla^2}{M^2}\right)R^{(1)\alpha\beta} + R^{(1)}\left(b+b'\frac{\nabla^2}{M^2}\right)R^{(1)}\right).
\end{equation}
We can extract from this the previous result for $R+R^2 /6M^2$ gravity by inserting $a=0$, $b=1/6$.
As before, we see a non-zero trace for a gravity theory with higher-order derivatives.

\subsection{Diffeomorphism invariance}\label{DiffInvSect}

Demonstrating diffeomorphism invariance is useful both as a consistency check of our derivation and to show that these results do not depend on any choice of coordinates. 
Let us apply the diffeomorphism transformation to the effective stress-energy tensor in (\ref{totalstress}).
The diffeomorphism transformation of a metric perturbation is given via the Lie derivative of the metric:
\begin{equation}
 \delta h_{\alpha\beta} = \mathsterling_\xi g_{\alpha\beta} = 2g_{\lambda(\alpha}\nabla_{\beta)}\xi^\lambda + g^{\gamma\delta}\xi_\gamma \nabla_\delta h_{\alpha\beta},
\end{equation}
where we have chosen to use the covariant derivative associated with the background metric $\nabla_\mu$, but we could have chosen to use a different covariant derivative.
Knowing that $h_{\alpha\beta}(\lambda)\sim\lambda$, we require that $\xi_\alpha(\lambda)\sim\lambda^2$ such that $\nabla_\alpha \xi_\beta (\lambda) \sim \lambda$.
Thus, in the weak limit, only the linearized diffeomorphisms are non-vanishing here. 
Before making use of the zero tensor, the result of varying (\ref{totalstress}) under diffeomorphisms is
\begin{eqnarray}\label{totaldifftrans}
 \kappa \delta t^{(0)}_{\mu\nu} & \w & -\xi_{(\mu}\nabla_{\nu)}R^{(1)}-\xi\cdot\nabla R^{(1)}_{\mu\nu} + 2\xi_{\alpha}\nabla_{(\mu}R^{(1)\alpha}_{\nu)}+\frac{1}{2}g^{(0)}_{\mu\nu}\xi\cdot\nabla R^{(1)} \nonumber \\ &&
+\frac{2}{M^2}\left(-\xi\cdot\nabla\nabla_\mu \nabla_\nu bR^{(1)} + 2\xi_{(\mu}\nabla_{\nu)}\nabla^2 bR^{(1)} - g^{(0)}_{\mu\nu}\xi\cdot\nabla\nabla^2 bR^{(1)}\right) \nonumber \\ &&
+\frac{1}{M^2}\left(-\xi\cdot\nabla\nabla^2 aR^{(1)}_{\mu\nu}+2\xi^{\alpha}\nabla_{(\mu}\nabla^2 a R^{(1)}_{\nu)\alpha}-\frac{1}{2}g^{(0)}_{\mu\nu}\xi\cdot\nabla\nabla^2 aR^{(1)}\right.\nonumber \\ &&
\left.-\xi\cdot\nabla\nabla_{\mu}\nabla_{\nu}aR^{(1)}+\xi_{(\mu}\nabla_{\nu)}\nabla^2 aR^{(1)}\right).
\end{eqnarray}
We take the form of the zero tensor given in (\ref{genzero}) and perform a diffeomorphism transformation to get
\begin{eqnarray}
 \xi_{(\rho}\nabla_{\sigma)}R^{(1)}_{\mu\nu} & \w & \frac{1}{M^2}\left(\xi_{(\rho}\nabla_{\sigma)}\nabla_\mu \nabla_\nu (a+2b)R^{(1)} + \frac{1}{2}g^{(0)}_{\mu\nu}\xi_{(\rho}\nabla_{\sigma)}\nabla^2 (a+2b)R^{(1)} \right.\nonumber \\ && 
\left.-\xi_{(\rho}\nabla_{\sigma)}\nabla^2 aR^{(1)}_{\mu\nu}\right).
\end{eqnarray}
The top line of (\ref{totaldifftrans}) has four terms at the third order in derivatives that we want to convert into higher-derivative expressions using the zero tensor.
The contracted forms of the zero tensor that we need are
\begin{equation}
 \xi_{(\mu}\nabla_{\nu)}R^{(1)} \w \frac{1}{M^2}\xi_{(\mu}\nabla_{\nu)}\nabla^2 (2a+6b)R^{(1)},
\end{equation}
\begin{equation}
 \xi\cdot\nabla R^{(1)}_{\mu\nu} \w \frac{1}{M^2}\left(\xi\cdot\nabla\nabla_\mu \nabla_\nu (a+2b)R^{(1)}+\frac{1}{2}g^{(0)}_{\mu\nu}\xi\cdot\nabla\nabla^2 (a+2b)R^{(1)}-\xi\cdot\nabla\nabla^2aR^{(1)}_{\mu\nu}\right), 
\end{equation}
\begin{eqnarray}
 -2\xi_{\alpha}\nabla_{(\mu}R^{(1)\alpha}_{\nu)} & \w & \frac{1}{M^2}\left(-2\xi\cdot\nabla \nabla_\mu \nabla_\nu (a+2b)R^{(1)}-\xi_{(\mu}\nabla_{\nu)}\nabla^2 (a+2b)R^{(1)} \right. \nonumber \\ && 
\left.+2\xi_{\alpha}\nabla_{(\mu}\nabla^2 aR_{\nu)}^{\ \ \alpha}\right),
\end{eqnarray}
\begin{equation}
 -\frac{1}{2}g^{(0)}_{\mu\nu}\xi\cdot\nabla R^{(1)} = -\frac{1}{M^2}g^{(0)}_{\mu\nu}\xi\cdot\nabla\nabla^2 (a+3b)R^{(1)}.
\end{equation}
Putting these terms together, we can see that, with the help of these zero tensor relations, the top line of (\ref{totaldifftrans}) cancels the rest of the terms exactly.
Thus the effective stress-energy tensor given in (\ref{totalstress}) is diffeomorphism-invariant.
This is an important check that gives us confidence in that result.

\section{Discussion and conclusion}\label{CONCL}

As reviewed in Section \ref{FRrev}, adding a Starobinsky $R^2$ term changes the result of the weak-limit calculation of the effective stress-energy tensor for backreaction such that the trace no longer vanishes in the weak limit.
This is true even though the background contribution of the $R^2$ term to the field equation vanishes in the weak limit. 
This tells us the $R^2$ term can still give a cosmologically important contribution to the backreaction after the universe has grown to a sufficient size that the $R^2$ term is no longer important in the pure background case.
If the $R^2$ term has a positive coefficient, $\kappa t^{(0)}$ converges to a negative value, as required for a candidate to mimic a positive cosmological constant.
Intutitively, this pure trace component can be attributed to there existing a ``scalaron'' mode in the $f(R)$ model that is most clearly apparent after performing a Legendre transformation into the equivalent scalar-tensor description.
This was the conclusion found in \cite{Preston:2014tua} for $R+R^2 /6M^2$ gravity.
This offered a motivation for alternative cosmological models that use exotic sources of inhomogeneity, such as the one discussed in \cite{Evans:2015zwa}. 

In this paper we have noted that this result is general for an $f(R)$ expansion that begins with $R+R^2 /6M^2$.
We then further generalized the procedure to find a general form for $\kappa t_{\mu\nu}^{(0)}$ in effective theories of gravity whose actions are expressible as a Taylor expansion in the Riemann tensor and covariant derivatives of the Riemann tensor, \ie local, manifestly diffeomorphism-invariant gravity theories.
The higher derivatives are balanced by powers of a mass scale for the effective theory, $M$.

We have argued that the physical consistency of the formalism requires us to scale this mass as $M\sim\lambda^{-1}$, as was also performed in \cite{Preston:2014tua}.
This is because the weak limit formalism describes the limit of a large averaging scale via $\lambda\to 0$, where $\lambda$ can be interpreted as a parameter proportional to the length scale for perturbations via (\ref{wavelength}).
Thus the derivative operator $\nabla_\mu/M$ effectively reads the ratio of the length scale of the theory to the perturbation length scale, which, being physical, should remain fixed as we tend the averaging scale to be large.
If we left $M$ as a constant in $\lambda$, not only would the ratio change, but the length scale of the perturbation would be driven below the cutoff scale of the effective field theory, which would be extremely pathological.
This scaling also ensures that we are comparing perturbation terms of the same order in $h$, such that $t^{(0)}_{\mu\nu}$ is written purely at $\mathcal{O}(h^2)$.
By this method, we wrote the complete set of action terms that contribute to $t_{\mu\nu}^{(0)}$ in closed form in (\ref{TotalAction}).
We derived the general form of $t_{\mu\nu}^{(0)}$ in closed form, as given in (\ref{totalstress}). Unlike in (\ref{StarStressTrace}), the new result in (\ref{totalstresstrace}) contains two independent structures whose coefficients have been left arbitrary in the general case.
Relating this to a specific example, we can look to the manifestly diffeomorphism-invariant classical Exact RG \cite{Morris:2016nda}, where we can specialize (\ref{totalstress}) to the effective action derived in the ``Einstein scheme'' simply by setting $a=-2b$.

For the choice of inhomogeneity model discussed in \cite{Evans:2015zwa}, the effective vacuum energy from backreaction was estimated to be of the form given in (\ref{EMSest}) in $R+R^2/6M^2$ gravity. The new result in (\ref{totalstresstrace}) has two new features. Firstly, it now has a potentially infinite expansion in higher-derivative operators, since $a$ and $b$ are both functions of $\nabla^2/M^2$. For each additional $\nabla^2/M^2$ operator found in a term, its additive contribution to $E_{\rm vac}^4$ would be suppressed by an extra factor of $(\Lambda_{\rm stripe}/M)^2$. Secondly, the trace now has two independent structures at each order in $M$, which have independent coefficients with unspecified signs. However, (\ref{EMSest}) would still be expected to be a sensible estimate for the magnitude of the backreaction in a generic case.

Although our effective stress-energy tensor has been derived with cosmological backreaction in mind, the method uses formalism that was originally constructed to describe gravitational waves from distant sources in \cite{Isaacson:1967zz,Isaacson:1968zza,Burnett:1989gp}.
This effective stress-energy tensor equally well applies to high-frequency gravitational waves.
Although it is necessary to introduce a background metric, we have not required any specific choice of background.
The only requirement imposed by the formalism for the matter content in the physical stress-energy tensor is that it satisfies the weak energy condition, as discussed in \cite{Green:2010qy}.
As demonstrated in Section \ref{DiffInvSect}, the effective stress-energy tensor in (\ref{totalstress}) is diffeomorphism-invariant.
Our new result includes the previous result as a special case where only the $R+R^2 /6M^2$ part is significant.
The non-zero trace found in (\ref{totalstresstrace}) motivates further research into the possible cosmological significance of backreaction in higher-derivative gravity models.
It also raises the possibility of using backreaction as a window on exotic ultraviolet physics that might otherwise be inaccessible.

\section*{Acknowledgements}

I acknowledge support from the University of Southampton through a Mayflower scholarship.
I thank Tim Morris for helpful comments.

\appendix
\section{Weak limits of individual field equation terms}\label{appindiv}
The weak limits of the individual field equation perturbations in Einstein gravity are
\begin{equation}
 \delta \left[R_{\mu\nu}\right] \w -\frac{1}{2}h^{\alpha\beta}R^{(1)}_{\mu\alpha\nu\beta} - \frac{1}{4}h_{\mu\nu}R^{(1)}+R^{(1)}_{\alpha(\mu}h_{\nu)}^{\ \ \alpha},
\end{equation}
\begin{equation}
 \delta\left[g_{\mu\nu}R\right] \w h_{\mu\nu}R^{(1)} - \frac{1}{4}g_{\mu\nu}^{(0)}\left(hR^{(1)}+2h^{\alpha\beta}R_{\alpha\beta}^{(1)}\right).
\end{equation}
For local $f(R)$ gravity, we must consider some additional field equation contributions:
\begin{equation}
 \delta\left[R_{\mu\nu}R\right]/M^2 \w R^{(1)}_{\mu\nu}R^{(1)}/M^2,
\end{equation}
\begin{equation}
 \delta\left[g_{\mu\nu}R^2\right]/M^2 \w g_{\mu\nu}^{(0)}R^{(1)2}/M^2,
\end{equation}
\begin{equation}
 \delta\left[D_\mu D_\nu R\right]/M^2 \w \frac{1}{2}\left(2R_{\mu\nu}^{(1)}+\nabla_\mu \nabla_\nu h\right)R^{(1)}/M^2,
\end{equation}
\begin{equation}
 \delta\left[g_{\mu\nu} D^2 R\right]/M^2 \w \left(\nabla^2 h_{\mu\nu}-\frac{1}{2}g_{\mu\nu}^{(0)}\nabla^2 h\right)R^{(1)}/M^2.
\end{equation}
The fully generalized gravity theory introduces new contributions again:
\begin{equation}
 \frac{1}{M^4}\delta\left[R_{\alpha\beta}D_{\mu}D_{\nu}a'\derarg R^{\alpha\beta}\right] \w \frac{1}{M^4}R_{\alpha\beta}^{(1)}\nabla_{\mu}\nabla_{\nu}a'\derbarg R^{(1)\alpha\beta},
\end{equation}
\begin{equation}
 \frac{1}{M^4}\delta\left[R D_{\mu}D_{\nu}b'\derarg R\right] \w \frac{1}{M^4}R^{(1)} \nabla_{\mu}\nabla_{\nu}b'\derbarg R^{(1)},
\end{equation}
\begin{equation}
 \frac{1}{M^2}\delta\left[g_{\mu\nu}R_{\alpha\beta}a\derarg R^{\alpha\beta}\right] \w \frac{g^{(0)}_{\mu\nu}}{M^2} R^{(1)}_{\alpha\beta}a\derbarg R^{(1)\alpha\beta},
\end{equation}
\begin{equation}
 \frac{1}{M^2}\delta\left[R_{\mu\alpha}a\derarg R_{\nu}^{\ \alpha}\right] \w \frac{1}{M^2}R^{(1)}_{\mu\alpha}a\derbarg R_{\nu}^{(1) \alpha},
\end{equation}
\begin{eqnarray}
 \frac{1}{M^2}\delta\left[D^2 a\derarg R_{\mu\nu}\right] & \w & \frac{1}{M^2}\left(-\frac{1}{2}ha\derbarg\nabla^2 R^{(1)}_{\mu\nu} + h_{\alpha(\mu}a\derbarg\nabla^2 R_{\nu)}^{(1)\alpha}\right. \nonumber \\ && 
\left.+ h^{\alpha\beta}a\derbarg\nabla_{\alpha}\nabla_{(\mu} R^{(1)}_{\nu)\beta}\right. \nonumber \\ && \left. -\frac{1}{2}h^{\alpha}_{\ (\mu}\nabla_{\nu)}\nabla_{\alpha} a\derbarg R^{(1)}\right),
\end{eqnarray}
\begin{eqnarray}
 \frac{1}{M^2}\delta\left[D_{\alpha}D_{(\mu}a\derarg R_{\nu)}^{\ \ \alpha}\right] & \w & \frac{1}{M^2}\left(\frac{1}{4}h_{\ (\mu}^{\alpha}\nabla_{\nu)}\nabla_{\alpha}a\derbarg R^{(1)} + \frac{1}{2}h^{\alpha\beta}a\derbarg \nabla_{\alpha}\nabla_{(\mu}R^{(1)}_{\nu)\beta}\right. \nonumber \\ && 
- \frac{1}{2}h^{\alpha}_{\ (\mu|}a\derbarg\nabla^2 R^{(1)}_{|\nu)\alpha} 
+\frac{1}{2}h^{\alpha\beta}a\derbarg\nabla_{\mu}\nabla_\nu R^{(1)}_{\alpha\beta} \nonumber \\ &&
\left.- \frac{1}{4}ha\derbarg\nabla_\mu\nabla_\nu R^{(1)}\right).
\end{eqnarray}
\section{Consistency of the trace of the effective stress-energy tensor}\label{apptrace}
There are two ways to evaluate the trace of the effective stress-energy tensor in weak limit, $t^{(0)}$.
Firstly, we can begin with the complete expression for $\kappa t^{(0)}_{\mu\nu}$, given in (\ref{totalstress}), and take the trace using the background metric.
Alternatively, we can take the trace of the field equation using the full metric first and then perform the weak limit to extract $t^{(0)}$.
To demonstrate that these two approaches give the same answer, consider the weak limit of the field equation perturbation:
\begin{equation}\label{generalstress}
 \kappa t^{(0)}_{\mu\nu} \w -\delta\left[\frac{2\kappa}{\sqrt{g}}\frac{\delta S_{\rm grav}}{\delta g^{\mu\nu}}\right].
\end{equation}
The zero tensor takes the form
\begin{equation}
 0 \w h_{\rho\sigma}\delta\left[\frac{2\kappa}{\sqrt{g}}\frac{\delta S_{\rm grav}}{\delta g^{\mu\nu}}\right].
\end{equation}
Pursuing the first approach, we can take the trace of (\ref{generalstress}) using the background metric to get
\begin{equation}
 \kappa t^{(0)} \w -g^{(0)\mu\nu}\delta\left[\frac{2\kappa}{\sqrt{g}}\frac{\delta S_{\rm grav}}{\delta g^{\mu\nu}}\right] \w -\delta\left[g^{\mu\nu}\frac{2\kappa}{\sqrt{g}}\frac{\delta S_{\rm grav}}{\delta g^{\mu\nu}}\right]-\underbrace{h^{\mu\nu}\delta\left[\frac{2\kappa}{\sqrt{g}}\frac{\delta S_{\rm grav}}{\delta g^{\mu\nu}}\right]}_{\text{zero, via the zero tensor}}.
\end{equation}
The right-hand side is the form for $\kappa t^{(0)}$ expected from the alternative approach of taking the trace of the field equation first and the weak limit second.
Thus both methods must give the same answer.

\bibliographystyle{hunsrt}
\bibliography{GeneralBackreact}

\end{document}